\documentstyle[preprint,aps]{revtex}

\begin{document}

\draft

\preprint{FIMAT-1/96}

\title{Can fractal-like spectra be experimentally observed in aperiodic
superlattices?}

\author{Enrique Maci\'a and Francisco Dom\'{\i}nguez-Adame}

\address{Departamento de F\'{\i}sica de Materiales,
Facultad de F\'{\i}sicas, Universidad Complutense,\\
E-28040 Madrid, Spain}

\maketitle

\begin{abstract}

We numerically investigate the effects of inhomogeneities in the energy
spectrum of aperiodic semiconductor superlattices, focusing our
attention on Thue-Morse and Fibonacci sequences.  In the absence of
disorder, the corresponding electronic spectra are self-similar.  The
presence of certain degree of randomness, due to imperfections occurring
during the growth processes, gives rise to a progressive loss of quantum
coherence, smearing out the finer details of the energy spectra
predicted for perfect aperiodic superlattices and spurring the onset of
electron localization.  However, depending on the degree of disorder
introduced, a critical size for the system exists, below which peculiar
transport properties, related to the pre-fractal nature of the energy
spectrum, may be measured.

\end{abstract}


\narrowtext

\section{Introduction}

One of the most appealing motivations for the experimental study of
aperiodic superlattices (ASLs) arranged according to Fibonacci
\cite{Merlin} and Thue-Morse \cite{Merlin2} sequences is the theoretical
prediction that these systems exhibit a highly-fragmented energy
spectrum displaying self-similar patterns \cite{KKT,Ryu,SST}. From a
strict mathematical perspective, it has been proven that the spectra of
both Fibonacci and Thue-Morse lattices are Cantor sets in the
thermodynamic limit \cite{Suto,Belli}. From an experimental point of
view, however, two major limitations appear to validate their peculiar
fractal nature.

In the first place, it is not possible to fabricate {\em perfect}
aperiodic structures.  Although X-ray diffraction studies show that the
characteristic structural order of ASLs is preserved under moderately
large growth fluctuations \cite{Todd}. the way this robust aperiodic
order can influence the transport properties of actual, defective ASLs
has not been received a proper treatment in the literature yet.  In this
sense, the observation of inhibition of vertical transport in periodic
superlattices with intentional disorder \cite{periodic}, in agreement
with the theory of localization in one-dimensional disordered systems,
suggests the possible existence of a competition between the long-range
aperiodic order and the unintentional short-range disorder and opens new
perspectives \cite{Liu}.

In the second place, even in the most favorable experimental
conditions, only {\em finite} arrangements with a limited number of
layers can be manufactured.  In this sense, the observation of
fragmentation patterns in the energy spectra of short ASLs using
different experimental techniques \cite{exper}, encourages further
analyses as to whether fractal-like spectra are to be expected in actual
ASLs with an increasingly {\em large} number of layers.

Two important questions then follow quite naturally: First, what are the
effects of unintentional disorder in the splitting scheme of the energy
spectrum of ASLs?  Second, what are the finite size effects on their
fractal-like properties?  In this paper we address these questions by
means of the study of the phase diagram, the localization length and the
bandwidth-scaling of the energy spectra.  Our results indicate that small
fluctuations in the sequential deposition of layers considerably smear
out the self-similarity of the energy spectra on increasing the system
size.  Thus, we conclude that fractal-like spectra with a richness of
finer details such those found by numerical analyses of {\em perfect}
ASLs are not to be expected in {\em large} ASLs, but quantum coherence
is strong enough in {\em short} systems to give rise to a measurable
hierarchical set of subminibands in the electronic spectra and to
influence its related transport properties.

\section{Model}

We consider quantum-well based GaAs-Ga$_{1-x}$Al$_x$As superlattices
with the same barrier thickness $b$ in the whole sample.  The height of
the barrier for electrons is given by the conduction-band offset at the
interfaces.  We take the origin of electron energies at the GaAs
conduction-band edge.  The thickness of each quantum-well is $\Delta
z_n-b\equiv z_n-z_{n-1}-b$, $z_n$ being the position of the center of
the {\em n\/}th barrier.  We will focus on electronic states close to
the bandgap and neglect nonparabolicity effects hereafter, so that a
one-band Hamiltonian suffices to describe those states.  Physical
magnitudes of interest can be easily computed using a transfer-matrix
formalism in this simple picture \cite{Rapid}.  In particular, we can
obtain subband energies under periodic boundary conditions and the
Lyapunov coefficient, which measures the inverse of the localization
length in units of the superlattice period.  We will consider Thue-Morse
superlattices (TMSLs) and Fibonacci superlattices (FSLs) since they have
been already constructed starting from two basic building blocks $A$ and
$A'$ by means of molecular beam epitaxy \cite{Merlin,Merlin2}.  In our
model, we take $A$ ($A'$) consisting of a quantum-well of thickness $a$
($a'$) and a barrier of thickness $b$.  ASLs are generated using the
following inflation rules: $A\rightarrow AA'$, $A'\rightarrow A'A$ for
the TMSL and $A\rightarrow AA'$, $A'\rightarrow A$ for the FSL. In this
way, finite and self-similar ASLs are obtained by $n$ successive
applications of these rules, with $N=2^n$ wells in the TMSL case and
$N=F_n$ wells in the FSL case.  The Fibonacci numbers are generated from
the recurrence law $F_n=F_{n-1}+F_{n-2}$, starting with $F_0=F_1=1$.

Interface roughness appears during growth in {\em actual} ASLs:
Protrusions of one semiconductor into the other cause in-plane disorder
and break translational invariance parallel to the layers.  Because the
in-plane average size of these defects depends on the growth conditions
and it is unknown in most cases, one is forced to develop a simple
approach.  We carry out such an approach describing local excess or
defect of monolayers by allowing $\Delta z_n$ to fluctuate uniformly
around the nominal values $a+b$ or $a'+b$.  This approach enables us to
restore the translational symmetry parallel to the layers, thus
facilitating computations.  Our approximation should be valid whenever
the mean-free-path of electrons is much smaller than the in-plane
average size of protrusions as electrons only {\em see} {\em
micro\/}-quantum-wells with small area and uniform thickness.
Therefore, each {\em micro\/}-quantum-well presents a slightly different
value of its thickness and, as a consequence, resonant coupling between
electronic states of neighbouring GaAs layers is decreased.  To get an
accurate description of electron dynamics, average over all possible
configurations of disorder is indeed required because the number of
interface defects as well as their mean thickness vary from layer to
layer.  For definiteness we take $\Delta z_n = a(1+W\epsilon_n)+b$ or
$\Delta z_n = a'(1+W\epsilon_n)+b$, where $W$ is a positive parameter
measuring the maximum fluctuation and $\epsilon_n$'s are distributed
according to a uniform probability distribution $P(\epsilon_n)=1$ if
$|\epsilon_n|<1/2$ and zero otherwise.  Note that $\epsilon_n$ is a
random {\em uncorrelated} variable, even when the lattice is constructed
with the constraint that the mean values of $\Delta z_n$ follow the
aperiodic sequences.

\section{Results and discussions}

We have studied GaAs-Ga$_{0.65}$Al$_{0.35}$As ASLs with $a=b=32\,$\AA\
and $a'=35\,$\AA.  In this case the conduction-band offset is
$250\,$meV and the effective masses are $m^*_{\mbox{\scriptsize
GaAs}}=0.067m$ and $m^*_{\mbox{\scriptsize GaAlAs}}=0.096m$, $m$ being
the free-electron mass.  In a periodic superlattice with
$a'=a=32\,$\AA\ only one allowed miniband lies below the barrier.  We
restrict ourselves to the study of the fragmentation of this miniband
when aperiodicity is introduced.  Averages over possible configurations
of disorder comprised a number of realizations varying from $50$ up to
$100$ to test the convergence of the computed mean values, and this
convergence was always satisfactory.  In order to compare with possible
experimental situations, we have considered $W$ ranging from $0$ up to a
maximum of $0.08$.  This value amounts to having maximum protrusion
thicknesses of half a monolayer on average.

Figure~\ref{fig1}(a) shows the dependence of the energy spectrum
structure on the amount of disorder, $W$, present in a FSL with
$N=F_{11}=144$ wells.  The energy is measured from the conduction-band
edge in GaAs.  For a perfect ($W=0$) FSL the overall structure of the
energy spectrum is characterized by the presence of four main subbands.
Inside each main subband the fragmentation pattern follows a
trifurcation scheme in which each subband further splits from one to
three subsubbands \cite{PRBFIBO,PRE}.  Therefore, the energy spectrum of
perfect and finite FSLs presents distinct pre-fractal signatures.  The
situation changes when randomness is introduced.  In fact, although the
tetrafurcation pattern of the perfect FSL still remain, the finer
details corresponding to successive steps in the hierarchical splitting
scheme are progressively smeared out on increasing the disorder due to
growth fluctuations.  Figure~\ref{fig1}(b) shows the dependence of the
spectrum structure with the degree of disorder present in a TMSL with
$N=128$ wells.  For a perfect system the fragmentation scheme agrees
with that previously discussed by Ryu {\em et al.\/} \cite{Ryu} in the
framework of the Kronig-Penney model, and displays pre-fractal
signatures as well.  The effects due to unintentional disorder are
similar to those shown in Fig.~\ref{fig1}(a) for a FSL. In a previous
paper we have demonstrated that pre-fractal signatures of FSLs can be
properly described by considering the resonant coupling between
electronic states of nearest-neighbor building blocks \cite{PRE}.  In
this sense, the above results reinforce this conclusion for they show
that the presence of the short-range disorder reduces the resonant
coupling between quantum-wells and, consequently, weaken the physical
mechanism giving rise to the self-similar pattern.

To investigate the effects due to the competition between long-range
aperiodic order and short-range disorder on the transport properties, we
have evaluated the localization length, $\ell$, as a function of the
energy, a magnitude which can be readily determined numerically within 
the transfer matriz formalism \cite{Rapid,Pendry}.
Figure~\ref{fig2}(a) compares the localization length for a FSL
with $N=F_{11}=144$ wells in two different cases.  For the perfect
($W=0$) case we obtain a very spiky structure.  Each peak corresponds to
a quasi-level, and most of them completely extends through the
superlattice, its localization length being one order of magnitude
greater than the system length.  Note that the distribution of peaks
reflects the overall fragmentation of the energy spectrum.  On the
contrary, when growth fluctuations are introduced, the localization
length distribution becomes smoother and its value remains {\em always}
smaller than the superlattice length, clearly revealing the onset of
localization effects.

In Fig.~\ref{fig2}(b) the same comparison is made for a TMSL with
$N=128$ wells.  Although the gross features of the plot, and their
corresponding physical interpretation, are completely analogous to the
case of the FSL, two interesting remarks are in order.  In the first
place, we observe that, in the perfect TMSL, there exists a
significant number of states whose localization length is several
orders of magnitude greater than the system size.  This fact is related
to the presence of latticelike states in the TMSL \cite{Ryu}.  The
occurrence of such states does not takes place in the FSL. In the second
place, the mean value of $\ell$ is lower for the defective TMSL, hence
indicating that the effects of localization are more intense in FSLs
than they are for TMSLs with the same amount of disorder.  A topic which
has deserved some attention recently concerns the comparison between the
transport properties in different kinds of aperiodically ordered
structures.  Our results indicate that, in presence of the same degree
of growth fluctuations, TMSLs should exhibit better transport properties
than FSLs.  Finally, notice that the localization onset is more
pronounced at the edges of the energy spectrum, meanwhile in the central
regions of the spectrum (about $110$ meV) the localization length almost
equals the system size.

From Figs.~\ref{fig1} and \ref{fig2} we can state that the general
effect of disorder in ASLs is twofold.  Firstly, it masks typical
pre-fractal features smearing out the hierarchical splitting scheme
taking place in perfect ASLs.  This effect is independent of the precise
nature of the underlying aperiodic order, this is to say, quasiperiodic
(FSL) or self-similar but not quasiperiodic (TMSL).  Secondly, the
overall decrease of the localization length of the electronic states
indicates that localization effects induced by random fluctuations are
starting on.

To get a more complete understanding of the competition between
long-range aperiodic order and random fluctuations, we have performed a
bandwidth-scaling study \cite{Koh}.  To this end, we define the
equivalent bandwidth $S$ as the sum of all allowed energy regions, thus
being nothing but the Lebesgue measure of the energy spectrum, and
compute it as a function of the number of wells $N$.  The obtained
results are presented in Fig.~\ref{fig3}.  Let us start discussing the
scaling of the Fibonacci system.  The equivalent bandwidth in the
perfect system [Fig.~\ref{fig3}(a)] decays as a power law of the form
$S(N) = S_0 N^{-\beta}$, where $S_0\sim 40\,$meV is close to the
bandwidth of the periodic superlattice and $\beta\sim 1/4$, within our
numerical accuracy.  According to earlier works \cite{Koh}, such a
behaviour is characteristic of a fractal-like energy spectra, becoming a
Cantor set in the limit $N \rightarrow \infty$.  However, in the
presence of moderate fluctuations ($W \leq 0.05$), the equivalent
bandwidth decays faster than a power law, as seen in Fig.~\ref{fig3}(b)
for $W=0.05$.  The deviation from a power law is a natural consequence
of the breaking of the self-similarity of the electronic spectrum
\cite{Koh}.  Using a least square fit we have found that $S(N) = S_0
N^{-\beta}\exp[-\alpha (W)N]$, where $\alpha(W)\sim 2W^2$.  From this
expression two consequences can be drawn.  First, the presence of the
exponential factor indicates that, when disorder is introduced in an
aperiodic system, its spectrum is not longer described in terms of a
{\em pure} singular continuous component, whose presence is
characterized by the presence of the power-law factor $N^{-\beta}$.
Second, the competition between aperiodic long-range order and local
disorder can be properly estimated from the deviation of $S$ from a pure
power-law behaviour.  Thus, localization effects dominate the electron
dynamics in ASLs whose length is larger than the {\em threshold} length
given by $N_{\text th}(W) \sim 1/\alpha(W)$.  This value is about
$N_{\text th}\simeq 200$ for $W=0.05$.  Interestingly the deviation from
power law decay becomes to be appreciable for moderately long systems,
so that short imperfect ASLs will show energy spectra quite similar to
that corresponding to a perfect ASL. This explains the success of
previous experimental works to detect self-similar features in actual
ASLs \cite{exper}.

Let us now consider bandwidth scaling in TMSLs.  As it was formerly
discussed by Riklund {\em et al.\/} \cite{Martin} the electronic
properties of Thue-Morse lattices lie between those of quasiperiodic
(Fibonacci) and usual periodic ones.  This point has been further
elaborated by Ryu {\em et al.\/} \cite{Ryu} who have shown that,
attending to its scaling behaviour, the spectrum of TMSL can be
decoupled into two components.  One component scales as $N^{-\beta}$
with $\beta \rightarrow 0^+$, and corresponds to wave functions with a
marked degree of extension (latticelike states).  The other one also
scales according to a power-law but $\beta>0$ in this case, and the
corresponding charge density distribution of the related wave functions
present large spatial fluctuations.  According to this, the scaling
behaviour shown in Fig.~\ref{fig3}(a) can be easily understood.  For
short TMSLs the contribution of both kinds of critical states to the
total bandwidth is essentially the same, for both spread uniformly over
the whole sample in an analogous manner.  But as the number of wells
progressively increases, the contribution to S due to the second kind of
wave functions decreases as a power law, whereas latticelike states give
an almost constant contribution to $S$.  As a consequence, only these
states contribute significantly to the value of $S$ in the thermodynamic
limit.  The maximum number of wells considered in our computations has
been $N=4096$, since the spectrum becomes too fragmented to compute
accurately $S$ for larger systems.  
Conversely, the equivalent bandwidth $S$ decays exponentially when
randomness takes place, as shown in Fig.~\ref{fig3}(b).  It is worth
mentioning that asymptotic values of $S$ are almost the same for both
kinds of disordered ASLs, indicating that the particular kind of
long-range order of the underlying structure is immaterial with regard
to their spatial extend.

\section{Conclusions}

In summary, we have proposed a realistic model to study {\em actual}
aperiodic superlattices by allowing the quantum-well thicknesses to
fluctuate around its nominal values in order to take into account
interface roughness.  Our results indicate that moderate fluctuations in
the sequential deposition of layers have significant effects on both the
energy spectrum and the spatial extension of wave functions.  The
fractal-like nature of an arbitrary spectrum is determined by two
complementary features.  In the first place, the energy spectrum becomes
more and more fragmented as the ASL length grows.  The physical origin
for this fragmentation stems from resonant tunneling effects between
electronic states of neighboring quantum-wells (short-range effects).
In the second place, the splitting scheme of the energy spectrum must
display a self-similar pattern.  The physical origin for this
self-similarity can be traced back to the structural self-similarity of
the superlattice itself which, in turn, is imposed by the aperiodic
ordering of the system (long-range effects).  Taking both facts into
account we conclude that the purported robustness of the aperiodic order
present in ASLs does not suffices, by its own, to guarantee the
fractal-like nature of the energy spectrum {\em in the presence of
disorder}, because of the main effect of growth fluctuations is
precisely to weaken the resonant coupling between electronic states.
Hence, albeit the structural aperiodic order is preserved in the
presence of moderate fluctuations, the self-similarity related to it
can not be properly expressed, in its finer details, in the energy
spectrum due to the loss of quantum coherence as a consequence of
short-range effects.

The relative importance that this competition between long-range order
and short-range disorder has on the transport properties depends
critically on the length of the system.  The values of the localization
length we have obtained indicate that wave functions no longer spread
over the whole ASL, as they do in the perfect case, but their degree of
extension amounts a significant fraction of the system size in
contrast with the usual view of localized states extending just over a
few wells. Thus, as a final conclusion, we can state that distinctive
features of a fractal-like spectrum can be experimentally observed in
ALSs of practical interest, whose length is smaller than a threshold
length. The value of this threshold length depends on the
heterostructure quality attained during the growth process.

\acknowledgments

Thanks are warmly due to Angel S\'anchez for helpful discussions and
critical reading of the manuscript. This work is supported by CICYT
through project MAT95-0325.



\begin{figure}
\caption{Phase diagram for a (a) GaAs-Ga$_{0.65}$Al$_{0.35}$As FSL with
$N=F_{11}=144$ wells and (b) GaAs-Ga$_{0.65}$Al$_{0.35}$As TMSL with
$N=128$ wells. Averages were taken over $10$ realizations of the
superlattice.}
\label{fig1}
\end{figure}

\begin{figure}
\caption{Localization length for (a) a perfect ($W=0$)
GaAs-Ga$_{0.65}$Al$_{0.35}$As FSL with $N=F_{11}=144$ wells (solid line)
as compared with a disordered FSL ($W=0.05$) of the same length (dashed
line), and (b) the same comparison for a GaAs-Ga$_{0.65}$Al$_{0.35}$As
TMSL with $N=128$ wells.  Averages were taken over $50$ realizations of
the superlattice.}
\label{fig2}
\end{figure}

\begin{figure}
\caption{Equivalent bandwidth $S$ as a function of the number of wells
in (a) perfect $W=0$ and (b) imperfect $W=0.05$
GaAs-Ga$_{0.65}$Al$_{0.35}$As FSL (solid lines) and TMSL (dashed lines).
Averages were taken over $50$ realizations of the superlattice.}
\label{fig3}
\end{figure}

\end{document}